\documentclass[preprint,12pt]{elsarticle}

\usepackage{amssymb}
\usepackage{amsmath}
\usepackage{subcaption}
\usepackage{graphicx}

\begin{document}

\begin{frontmatter}



\title{Existence of dynamical fluctuation in AMPT generated data for Au+Au collisions at 10 AGeV} 



\author{Somen Gope$^{*}$}
\author{Supriya Das}
\author{Saikat Biswas}
\address{Department of Physical Sciences, Bose Institute Unified Academic Campus, Kolkata, India \\
$^{*}$somengope30@jcbose.ac.in
}

\begin{abstract}

In this study, the intermittency behavior of emitted particles produced in heavy ion collisions has been studied using both modes (default $\&$ string melting) of A Multi Phase Transport (AMPT) model-generated data. We adopted one of the most conventional and successful techniques, the Scaled Factorial Moment (SFM) method, using Monte Carlo (MC) data for 10 AGeV Au+Au collisions in search of intermittency in the model-generated data. Our interest is to search for intermittency behavior of particles that leads to multiplicity fluctuations and that would reveal a phase transition from hadronic matter to QGP. In this article, the intermittency values for both modes of AMPT data are presented. The results obtain some insight into the dynamics of heavy ion collisions and the formation of QGP.


\end{abstract}




\end{frontmatter}

\section{Introduction}
\label{sec:1}

One of the primary objectives of relativistic heavy-ion collisions is to investigate the QCD phase diagram and phase boundary \cite{n_p1,n_p2,n_p3}. Finite temperature Lattice QCD simulations indicate a crossover from hadronic phase to Quark Gluon Plasma (QGP) phase at vanishing baryon chemical potential $\mu_{B}=0$ and a QCD-based model \textit{``Two Color QCD"} suggests that the transition might be first-order at large $\mu_{B}$ \cite{n_p4,n_p5,p1,p111}. It is known as the critical end-point (CEP) at which the first-order phase transition terminates. Theoretically and experimentally, efforts are being made to locate the CEP by studying the non-monotonic behaviors in the fluctuations of conserved quantities. Multiplicity or density fluctuations that follow intermittency behavior arise from the linked fractal geometry of QCD matter, in analogy to the critical opalescence seen in conventional matter around critical point. Rigorous studies have already been done by various researchers and research groups using experimental data. Power-law behavior in $Si+Si$ collisions at 158 AGeV has been seen in the NA49 experiment \cite{p1111}. The scaling exponent derived from the intermittency index in the initial analysis of the STAR experiment indicates a minimum in central Au+Au collisions at $\sqrt{S_{NN}}$ = 20-30 GeV \cite{p11111}. Meanwhile, simulations involving hadronic potentials and the self-similar feature indicate that the intermittency behavior could be seen in $Au+Au$ collisions at RHIC energy. Investigating the same for lower beam energy is also crucial. According to hydrodynamical calculations, it is expected that there might be a first-order phase transition from hadronic matter to exotic nuclear matter at the highest SIS100 energy \cite{p2}. It is also anticipated that a critical point may exist in such a region of QCD phase diagram. Around the critical point, the thermodynamical variables are extremely sensitive to slight changes in the order parameter. Searching for physical observables related to thermodynamical parameters might be important to study. \par

Multiplicity fluctuation is one of the important observables to provide the information about particle production mechanism in heavy-ion collisions. The fluctuation of multiplicity affects intermittency, and there is a correlation between these two. Intermittency measurement in relativistic heavy-ion collisions is a helpful probe of density fluctuations associated with the events. The Scaled Factorial Moment (SFM) technique is one of the most successful methods and is used to extract dynamical fluctuations from a mixture of statistical and dynamical fluctuations. In this study, an attempt has been made to study the dynamical fluctuation with the SFM technique using the default mode of a multi-phase transport model (AMPT) \cite{p_ampt_01,p_ampt_02,p_ampt_03,p_ampt_04} generated data.

This work aims to analyze AMPT-generated data using the scaled factorial moment technique to detect any presence of intermittency in the data and to determine if string melting influences the intermittent particle emission observed in nuclear collisions. It is expected that, according to hydrodynamical calculations, the deconfinement phase border is first expected to be reached at the highest achievable energy of SIS100, FAIR \cite{fair_energy_01,fair_energy_02}. Unlike the RHIC at BNL, USA, and the Large Hadron Collider (LHC) at CERN, Switzerland, the FAIR-CBM experiment is expected to generate a fireball with a high net baryon density. In this context, the final state interactions between the produced particles in the collision can have a significant impact on the observables of these collisions.

\section{Model Description}
A multi-phase transport (AMPT) is one of the most successful Monte Carlo transport models for heavy ion collision in a large energy range of energies. It consists of four major parts:

\begin{itemize}
	\item Initial condition.
	\item Interaction between partons.
	\item Hardronization: Lund string fragmentation model used for default AMPT.
	\item Production of final particle spectra.
\end{itemize}

There are various successful MC models, such as transport models, hydrodynamic models, and hybrid models, that have been developed to simulate the phase space evolution of matter produced in high-energy heavy-ion collisions \cite{p_model_01}. The multi-phase transport (AMPT) model is one such example. This model employs a kinetic theory approach to explain the progression of heavy-ion collisions.\par

The initial state of AMPT relies on the dual-component model of the Heavy Ion Jet Interaction Generator (HIJING); this includes the momentum and spatial distribution of minijet partons and string excitation. Two-body scattering is considered for Zhang's parton cascade (ZPC) mechanism and the cross-section derived from pQCD with screening masses.

\subsection{Default AMPT}

The default version of the AMPT model was first developed. After the initialization, the minijet partons enter the parton cascade and eventually recombine with their parent strings to form hadrons via the Lund string fragmentation. The fragmentation function is mathematically defined as:

\begin{equation}
f(z) \propto \frac{(1-z)^{a}}{z} exp(-bm_{T}^2/z)
\end{equation}

where $z$: light-cone momentum fraction of the produced hadrons with respect to that of the fragmentation string.\\
$m_{T}$: transverse mass; and
$a$, $b$ are free parameters.  \\

The default AMPT model performs well to reproduce the rapidity distributions and transverse momentum spectra of identified particles produced in heavy ion collisions at SPS and RHIC energies. However, it notably fails to accurately predict the elliptic flow ($v_{2}$) observed specifically at RHIC.

\subsection{AMPT with String Melting}

The partonic phase in AMPT with string melting mode is defined alternatively. String melting in its place replaces the conventional hadronization process where the strings between partons are fractioned into hadrons. In this case the partons created by parton strings in the initial stage of the collision are allowed 'to melt' into partons. This is a phase transition: melting strings into a quark-gluon plasma (QGP) phase. \par

In summary, the AMPT default mode focuses more on hadronic interactions with a limited partonic phase, while the string melting mode assumes a full partonic phase and is better suited for heavy ion collisions where a quark-gluon plasma is expected to form. Also, string melting mode of AMPT might be suitable for strongly interacting matter under extreme matter density condition in HIC. 

\section{Theoretical formulation}
\label{sec:2}

In the intermittency study, the observable of interest is the scaled factorial moment of particle multiplicity distribution. It depends on the number of subdivisions of phase space (say rapidity, azimuthal angle, momentum).

In one-dimensional phase space, the interval $\Delta x$ is divided into $M$ bins, each with a width $\delta x = \frac{\Delta x}{M}$. It is essential for exploring particle distributions in the phase space more effectively. By observing how particles are distributed across these bins, researchers can probe the underlying dynamics of the system to reveal patterns and correlations.

Let us consider that $K_{m}$ represents the number of particles in the $m^{th}$ bin, where $m$ varies from $1$ to $M$. The distribution of these particles across the bins is not arbitrary; it reflects the complexity of interactions that uncover the behavior of the system. To study such complexity, one can use the most successful technique by the name of factorial moment, a powerful mathematical tool used to extract the dynamical fluctuations from the mixture of dynamical $\&$ statistical fluctuation within the system. The factorial moment of order $q$, denoted as $f_q$, is defined as follows:

\begin{equation}
f_{q} = \langle K_{m}(K_{m}-1).....(K_{m}-q+1) \rangle
\end{equation}

This simple equation holds significant meaning. The factorial moment $f_q$ shows how particles in a bin are related, giving insight into the microscopic processes affecting the system \cite{mf_p1}. We take the average, which can be done in two different ways to reveal different aspects of the fluctuations:

\textit{Vertical Averaging:} In this approach, the $f_{q}$ is averaged over all events for a fixed bin. This method reveals the event-by-event fluctuations, providing a macroscopic view of how the distribution of particles varies for different events. 

\textit{Horizontal Averaging:} Conversely, horizontal averaging focuses on a single event, averaging the $f_{q}$ across all bins within that event. This approach highlights the spatial fluctuations within the phase space of an event, offering a detailed picture of how particles are distributed across the bins.

In the analysis of particle distributions, the statistical component of these fluctuations follows a Poisson distribution, a common model for random, uncorrelated events. However, in reality, the system often exhibits dynamics that exhibit more complexity than randomness \cite{mf_p2}. Bialas and Peschanski demonstrated that the factorial moments of the multiplicity distribution across a large sample of events are sensitive to dynamical fluctuations, effectively separating them from the statistical one \cite{paper4}. They showed that if the probability distribution $P_n$ of $K_{m}$ can be expressed as a convolution of a dynamical distribution $D(\nu)$ and a Poissonian statistical distribution, then the factorial moment $f_q$ acts as a filter, removing the statistical fluctuations and leaving behind a pure signal \cite{paper2111}.

To further explore the dynamics of a single event, the $q^{th}$ order scaled factorial moment, defined as \cite{mf_p3}:

\begin{equation}
F_q = M^{q-1} \sum_{m=1}^{M} \frac{K_{m} (K_{m} - 1) \dots (K_{m} - q + 1)}{K(K - 1) \dots (K - q + 1)}
\end{equation}

In this equation, $K$ represents the total multiplicity of the event, calculated as $K = \sum_{m=1}^{M} K_{m}$. 

When the analysis extends to a large number of events, each with varying multiplicities, the expression for the scaled factorial moment modifies. The modified form of SFM is given by the following equation:

\begin{equation}
F_q = M^{q-1} \sum_{m=1}^{M} \frac{K_{m} (K_{m} - 1) \dots (K_{m} - q + 1)}{\langle K \rangle^q}
\end{equation}

Here, the average multiplicity $\langle K \rangle$ is defined as $\langle K \rangle = \frac{\sum_{1}^{N_{ev}} K}{N_{ev}}$, where $N_{ev}$ is the total number of events.

And the horizontally averaged scaled factorial moment expressed as:

\begin{equation}
\langle F_q \rangle = \frac{1}{N_{ev}} \sum_{i=1}^{N_{ev}} M^{q-1} \sum_{m=1}^{M} \frac{K_{m} (K_{m} - 1) \dots (K_{m} - q + 1)}{\langle n \rangle^q}
\end{equation}

The horizontally averaged SFM provides a comprehensive measure of the overall fluctuation pattern across the sample of events. By examining how $\langle F_q \rangle$ behaves as the bin width $\delta x$ one can identify the presence of the intermittency in the particle emission in high-energy collisions. In a log-log plot, a linear relationship between $\langle F_q \rangle$ and $M$ signifies a power-law behavior, described by $\langle F_q \rangle \propto M^{\alpha_q}$. This power-law behavior is a signature of intermittency, indicating that the particle emission process exhibits a self-similar, fractal-like structure across different scales.

The concept of intermittency is deeply connected to the ideas of self-similarity and fractal geometry, both of which are foundational to our understanding of complex systems. In systems that exhibit intermittency, the distribution of particles follows a fractal pattern, where similar structures are observed at different levels of magnification. This fractal behavior is characterized by the anomalous fractal dimension $d_q$, which is related to the intermittency index $\alpha_q$ through the equation:

\begin{equation}
d_q = \frac{\alpha_q}{q - 1}
\end{equation}

The fractal dimension $d_q$ serves as a quantitative measure of the complexity of the particle distribution, offering insights into the nature of the underlying dynamics. By studying the dependence of $d_q$ on the order $q$, one can obtain valuable information about the mechanisms of particle production \cite{mf_p4}. An increase in $d_q$ with $q$ suggests that the particles are produced via some branching mechanism \cite{mf_p5}.

On the other hand, if $d_q$ remains unchanged across different $q's$, it indicates a phase transition. Such transitions often involve a change in the state of the system, where fluctuations at all scales contribute equally to the dynamics, leading to a uniform fractal dimension with $q$.

\section{Results}
\label{sec:3}

The analysis was with the generation of AMPT events using both the default and string melting modes. To check the model’s effectiveness at FAIR energies, a comparison was made with experimental data from the E895 experiment \cite{Experimental_data_p1}.

\begin{figure}[ht!]
	\centering
	\begin{subfigure}[b]{0.5\textwidth}
		\centering
		\includegraphics[width=\textwidth]{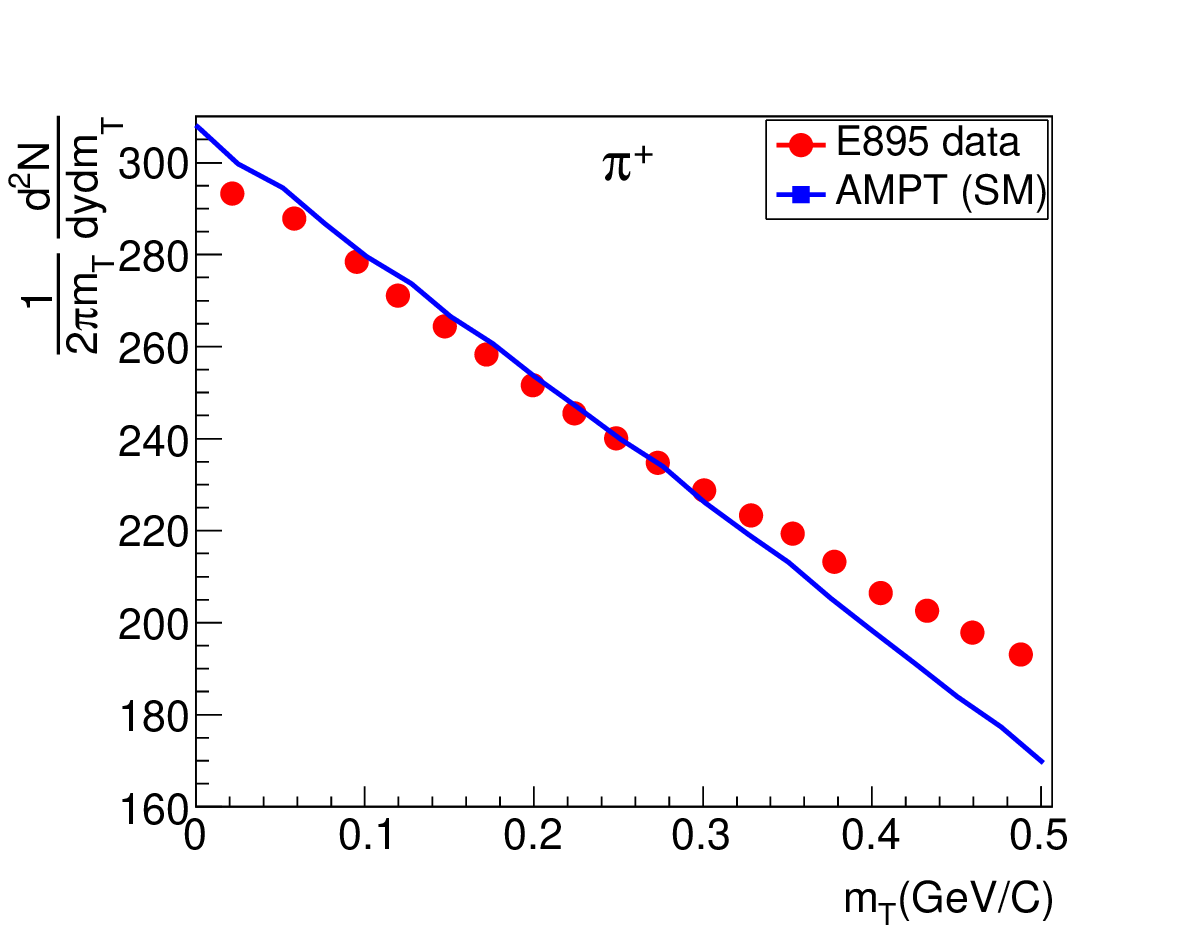}
		\caption{}
		\label{fig:y equals x}
	\end{subfigure}
    \hskip -2ex
	\begin{subfigure}[b]{0.5\textwidth}
		\centering
		\includegraphics[width=\textwidth]{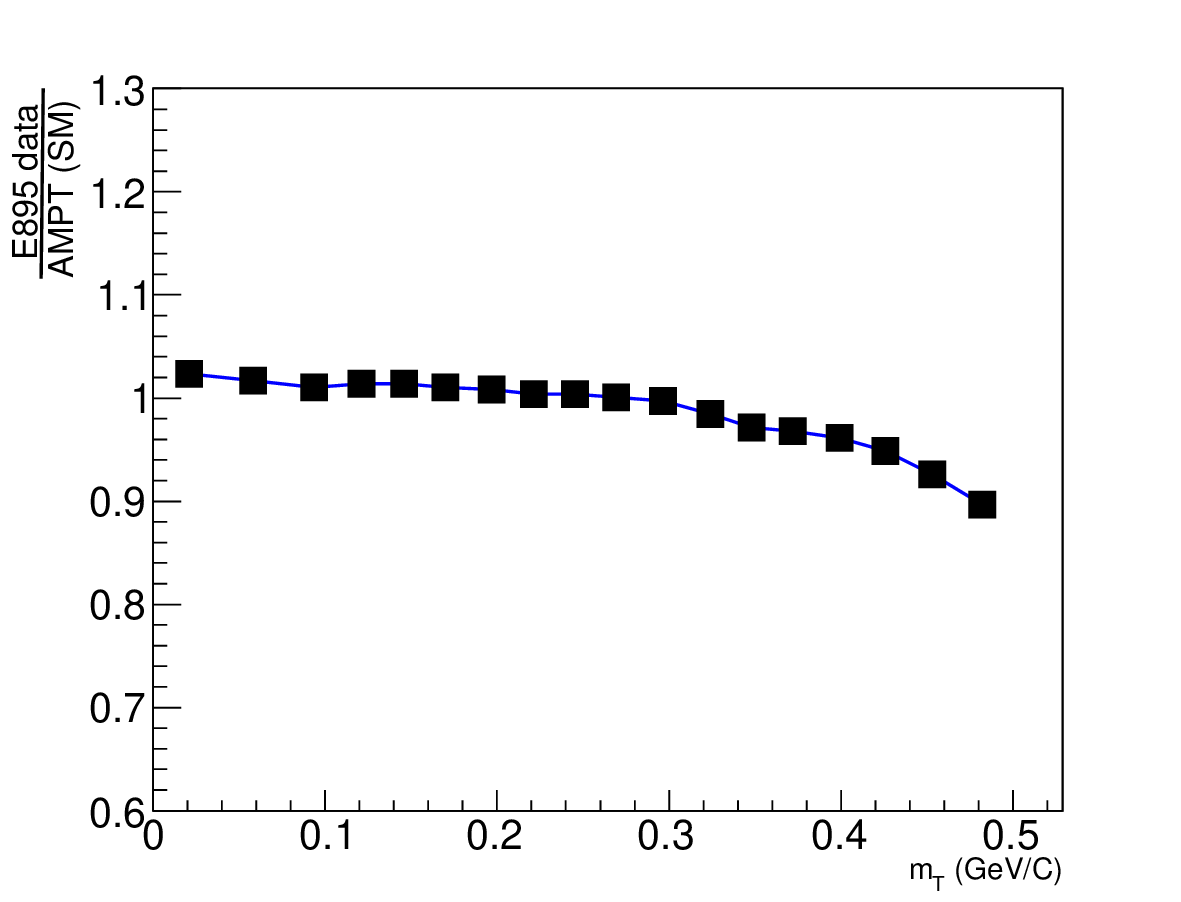}
		\caption{}
		\label{fig:three sin x}
	\end{subfigure}
	\caption{$m_{T}$ spectra of $\pi^{+}$ at 8 $AGeV$ energy using (a) AMPT generated data, and (b) ratio of $m_{T}$ spectra of experimental (E895 experiment\cite{Experimental_data_p1}) data to AMPT generated data}
\end{figure}

Plotted the $m_{T}$-spectra of $\pi^{+}$ shown in Fig. 1(a) using AMPT (SM) generated data, which is represented by a blue solid line, and experimental data with red markers. The comparison involved taking the ratio of transverse mass $m_{T}$ spectra between the E895 data and the AMPT-generated MC data.

The plot shows that the AMPT (SM) model successfully reproduces the $m_{T}$ spectra of the E895 experiment up to approximately 0.3 $GeV/c$. Beyond this, the values of $m_{T}$ overestimate, and there is a significant deviation from experimental data. This trend suggests that the AMPT model is quite effective for describing data at lower energies. 

\begin{figure}[ht!]
	\centering
	\includegraphics[width=.80\linewidth]{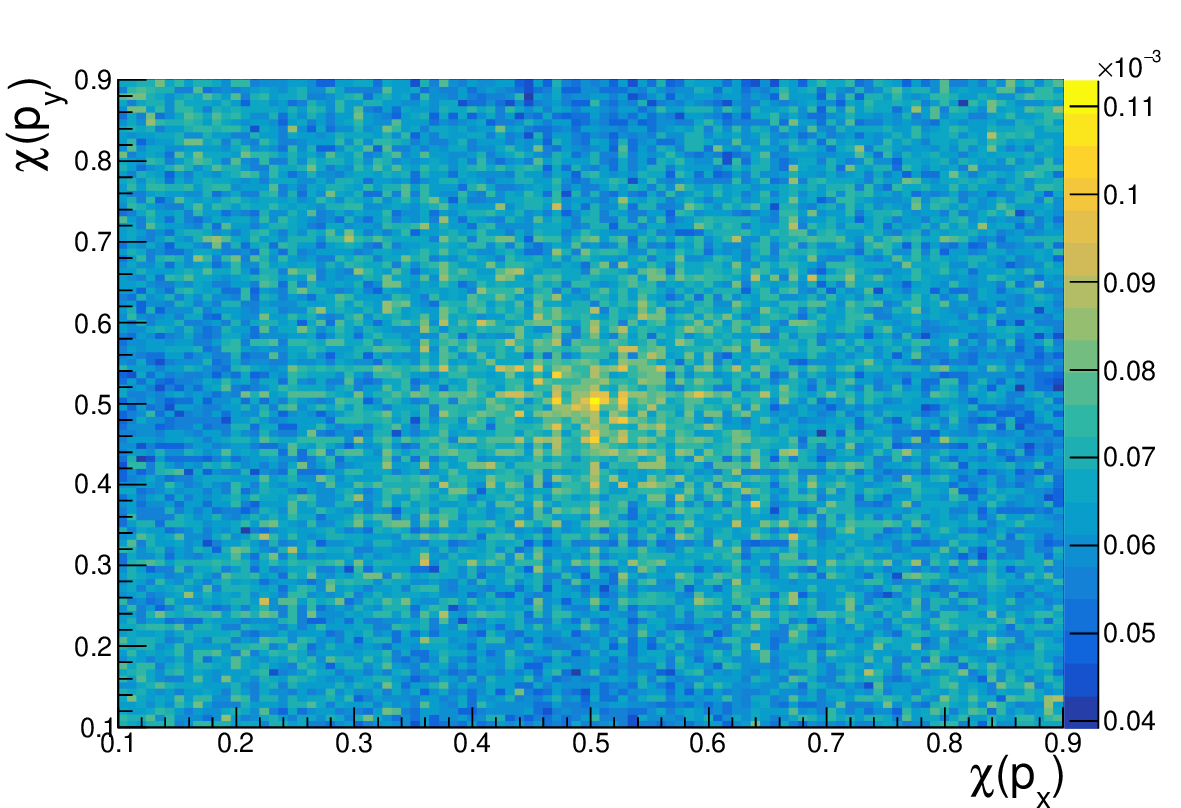}  
	\caption{$\chi(p_{x}) vs \chi(p_{y})$ distribution for charged particles produced in 10 AGeV Au+Au collisions using AMPT event generator.}
	\label{fig:sub-first}
\end{figure}

To reduce any potential projection effects, the analysis has been carried out for all charged particles in the two-dimensional $p_{x}-p_{y}$ space. To eliminate the initial shape dependence of the two-dimensional density distribution, the $p_{x}$ and $p_{y}$ values of each primary charged particle are converted into new variables, $\chi(p_{x})$ and $\chi(p_{y})$. These transformations are defined as follows:

Replace $p_{x}$ by $p_{y}$ in the following equation for $\chi(p_{y})$.

\begin{figure}[ht!]
	\centering
	\includegraphics[width=0.8\linewidth]{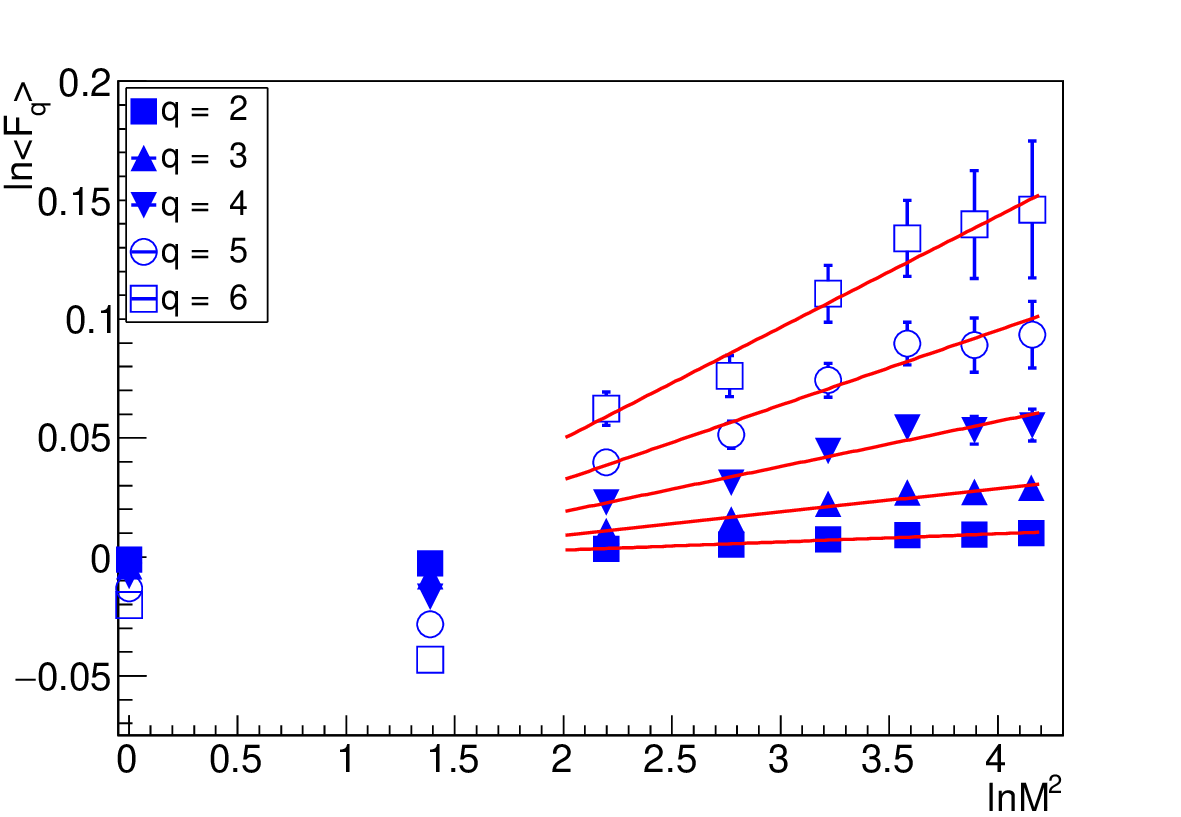}  
	\caption{Variations of $\langle F_q \rangle$ against $lnM^{2}$ for charged particles produced in Au+Au collisions at 10 AGeV energy using AMPT generator with default mode.}
	\label{fig:sub-first}
\end{figure}

\begin{figure}[ht!]
	\centering
	\includegraphics[width=0.8\linewidth]{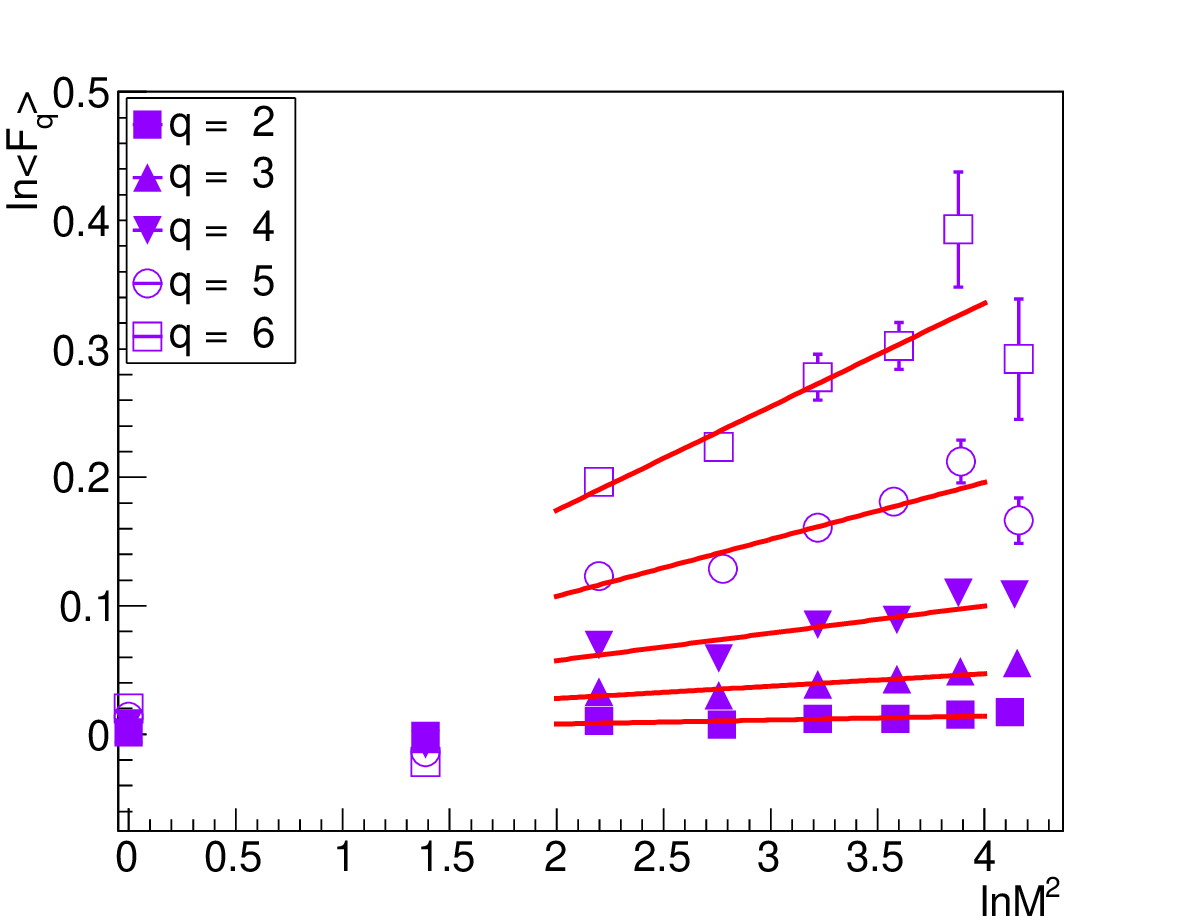}  
	\caption{Variations of $\langle F_q \rangle$ against $lnM^{2}$ for charged particles produced in Au+Au collisions at 10 AGeV energy using AMPT generator with string melting mode.}
	\label{fig:sub-first}
\end{figure}

where $p_{x,min}$ and $p_{y,min}$ are the minimum values of $p_{x}$ and $p_{y}$, respectively. $p_{x,max}$ and $p_{y,max}$ are the maximum values of $p_{x}$ and $p_{y}$. The values of $\chi$ will lie in between $0$ and $1$. Other parameters are defined in one of our earlier published papers \cite{p1}.\\

Now, plotted the new cumulant variable for $p_{x}$ and $p_{y}$, which is shown by Fig. 2). The figure shows a 2D $\chi(p_{x})-\chi(p_{y})$ distribution of particles that appears uniform and flat, and the flatness indicates that the distribution is free from any preferential emission patterns. Thus effectively reduces error in the dynamical fluctuation studies. This approach reduces any influence from the initial kinematic shape dependence of the single particle spectra and ensures precise analyses of the fluctuations.

\begin{figure}[ht!]
	\centering
	\includegraphics[width=.8\linewidth]{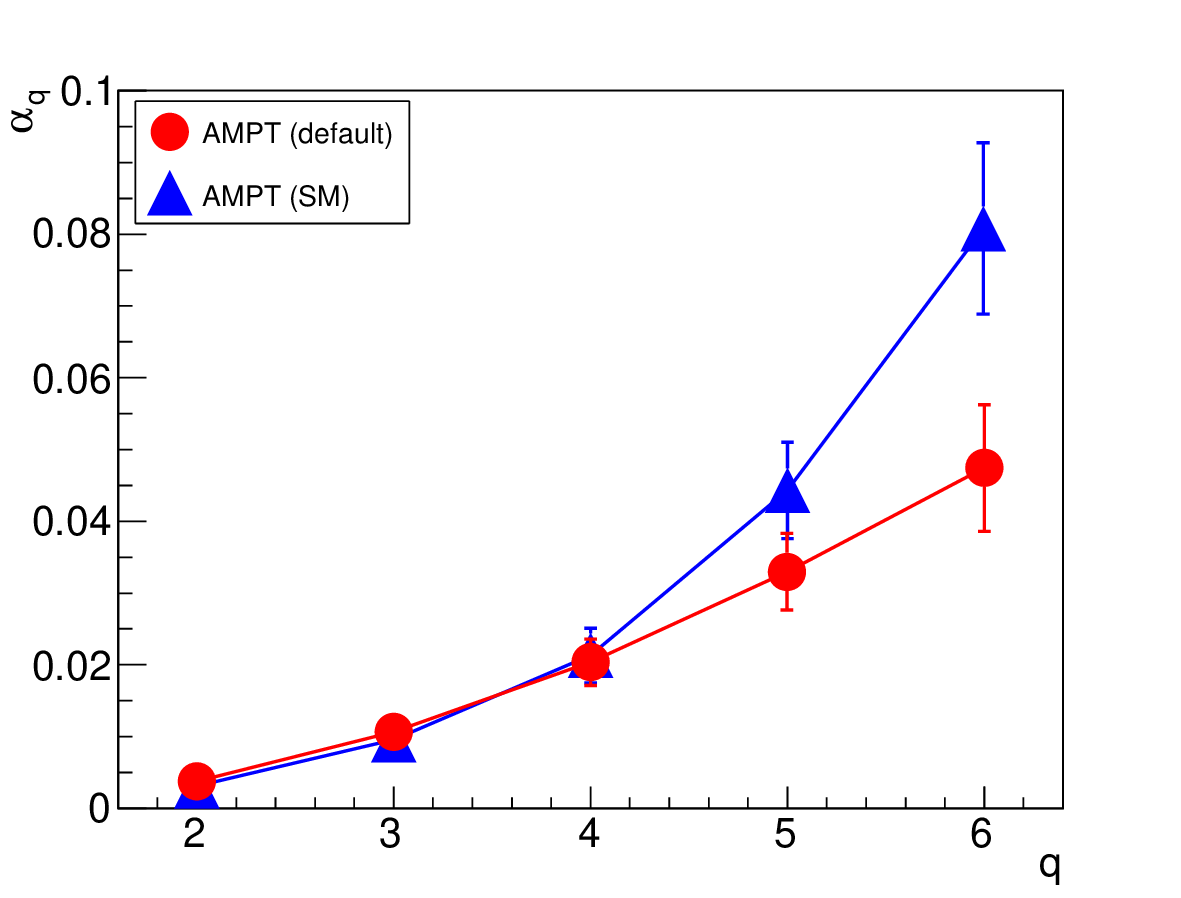}  
	\caption{$\alpha_{q}$ vs. $q$ plot for charged particle produced uing both the modes of AMPT generator.}
	\label{fig:sub-first}
\end{figure}

\begin{figure}[ht!]
	\centering
	\includegraphics[width=0.8\linewidth]{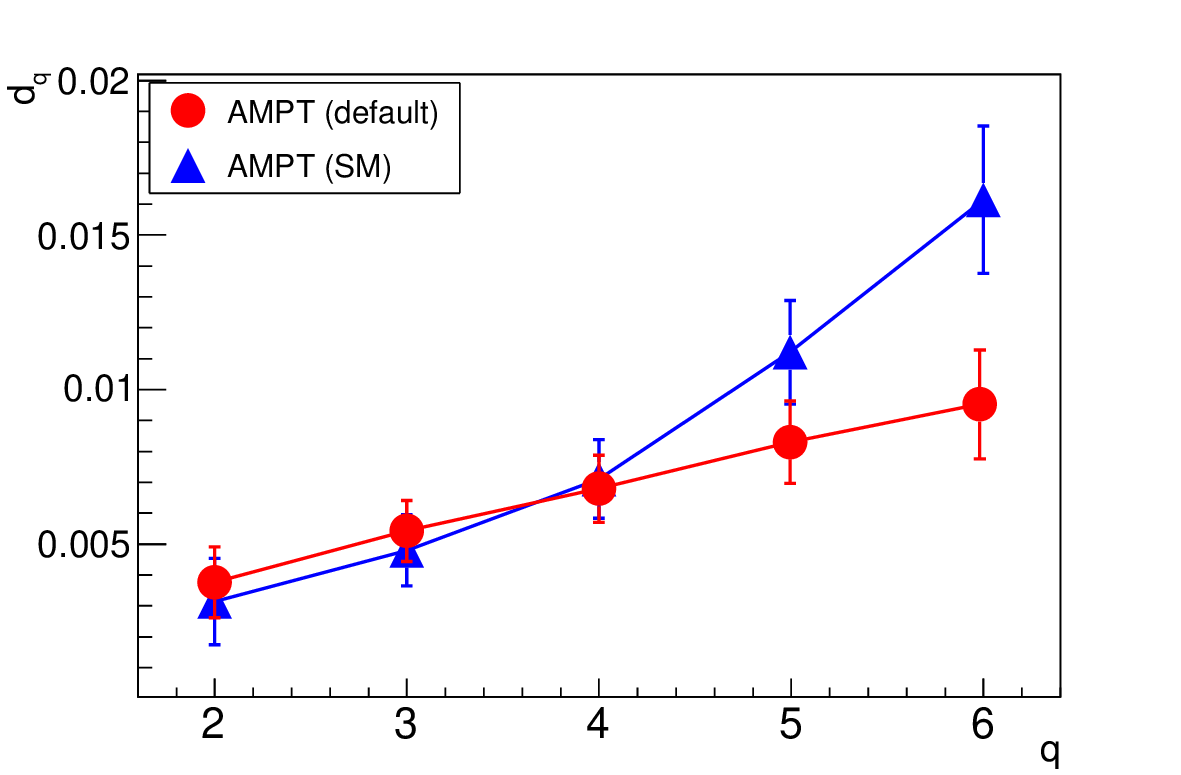}  
	\caption{The variation of $d_{q}$ with $q$ plot for all cahrged particles produced in Au+Au collisions at 10 AGeV energy using both modes of AMPT generator.}
	\label{fig:sub-first}
\end{figure}

In this study, the SFM is derived in the two-dimensional $\chi(p_{x})-\chi(p_{y})$ space and undertaken a rigorous and systematic analysis as described by the formula in Eqn 4. Analysis was started by partitioning the $\chi(p_{x})-\chi(p_{y})$ space into bins, where the total number of bins is calculated by $M_{i} \times M_{i} = M^{2}$. The sizes of the bins follow the sequence of 1, 4, 9, 16, 25, 36, and so on up to 100, and the uniform width of each bin is $d\chi(p_{x}) \times d\chi(p_{y})$.


To compute SFM for a bin, we carefully count the number of particles in each bin and use these counts for that particular bin. The SFMs of each bin are averaged across all the bins for the given configuration, and it provides the measure of dynamical fluctuation. To calculate the $\langle F_q \rangle$ for different values of $M^{2}$ the mean value of SFMs is combined for all the events.

The present study provides profound insights into the underlying statistical behavior and dynamical characteristics of the system, enriching our understanding of its fundamental properties.

\begin{table*}[ht!]
	\caption{The table of intermittency indices for 10 AGeV Au+Au data using both modes of AMPT data.}
	\label{sphericcase}
	\begin{tabular*}{\textwidth}{@{\extracolsep{\fill}}lrrrrl@{}}
		\hline
		Systems & \multicolumn{1}{c}{$\alpha_{2}\times10^{-3}$} & \multicolumn{1}{c}{$\alpha_{3}\times10^{-3}$} & \multicolumn{1}{c}{$\alpha_{4}\times10^{-3}$} & \multicolumn{1}{c}{$\alpha_{5}\times10^{-3}$} & $\alpha_{6}\times10^{-3}$ \\
		\hline
		AMPT  & 3.80 $\pm$ 1.10 & 10.7$\pm$ 1.90 & 20.4$\pm$ 3.20 & 32.8$\pm$ 5.30 & 47.4$\pm$ 8.80 \\
		(default) &&&&& \\
		\hline
		AMPT  & 3.10$\pm$ 1.40  & 9.60$\pm$ 2.30  & 21.3$\pm$ 3.80  & 44.3 $\pm$ 6.70 & 80.7$\pm$ 11.9 \\
		(SM)&&&&& \\
		\hline
	\end{tabular*}
\end{table*}
The $\langle F_q \rangle$ for different values of $q$ are then calculated for $\chi(p_{x}-p_{y})$ space using both the sets of AMPT (default) and AMPT (SM) generated data. The $\langle F_q \rangle$ vs. $lnM^{2}$ are then plotted and are shown in Fig 3 and Fig 4, respectively. The Fig 3 \& Fig 4 imply a clear indication of power-law behavior with the form $\langle F_q \rangle \propto M^{\alpha_{q}}$. This behavior confirms the presence of intermittency in both modes of AMPT-generated data. The error estimation is one of the most important tasks in the calculation of $\langle F_q \rangle$. The errors in this analysis are only independent statistical errors not considered for different bin sizes. Several other workers reported in their work that exclusion of bin size-dependent error does not change the outcomes appreciably \cite{paper6666,paper66666,paper666666}.

From this study it is observed that the values of $\langle F_q \rangle$ rise with $lnM^{2}$ values for both the data of AMPT. This trend is unvarying across both the default and string melting (SM) modes of the AMPT model. The values that represent this behavior are listed in Table 1; the intermittency indices for various moment orders have been meticulously calculated from different datasets. Of particular note, the intermittency indices for $q=5$ and $q=6$ are markedly higher in the AMPT (SM) mode compared to those in the AMPT (default) mode, indicating a more pronounced fluctuation strength in the string melting scenario.

\begin{figure}[ht!]
	\centering
	\includegraphics[width=0.8\linewidth]{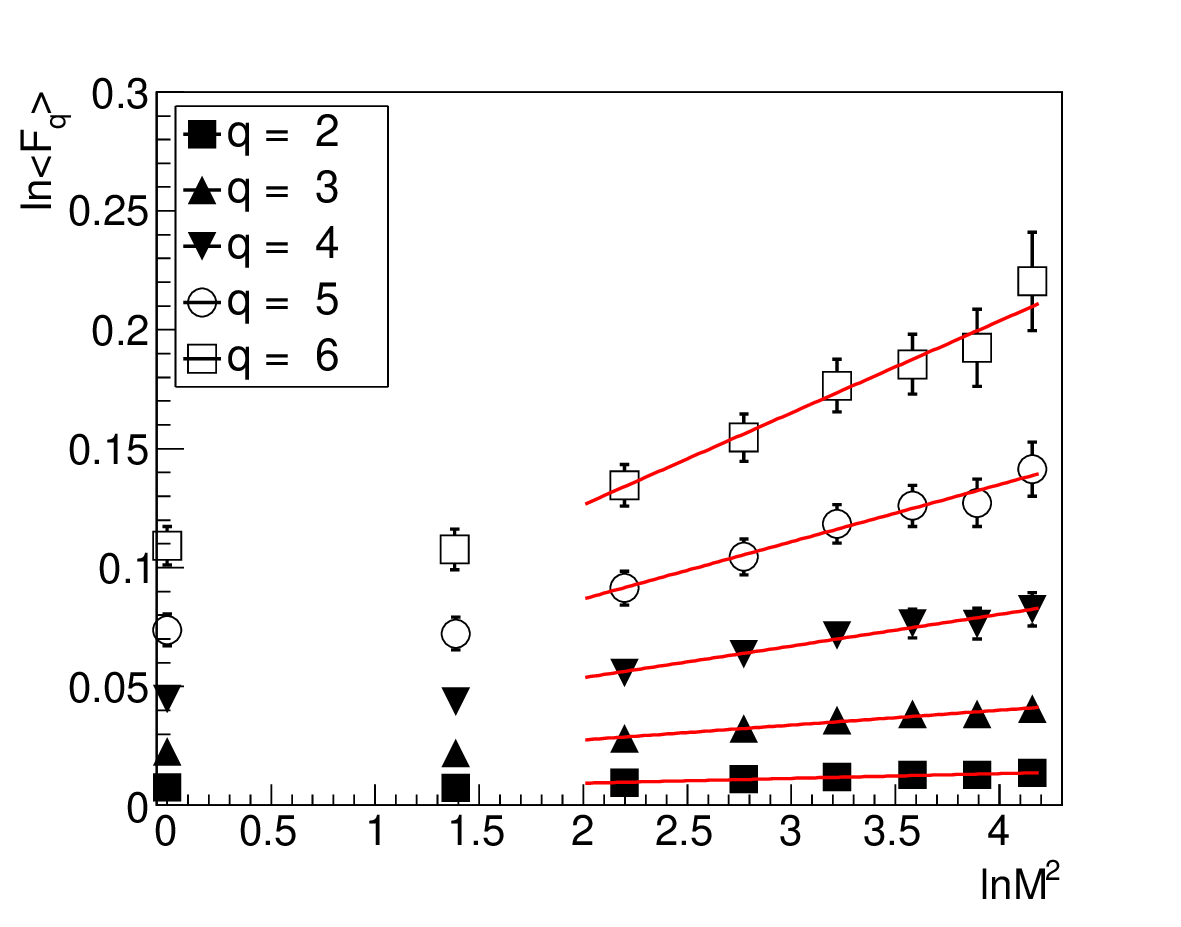}  
	\caption{The $\langle F_q \rangle$ vs. $lnM^{2}$ plots for all charged particles produced in Au+Au collisions at 200 GeV energy using AMPT(default) generator.}
	\label{fig:sub-first}
\end{figure}

Fig. 6 shows the variation of anomalous fractal dimension $d_{q}$ with order of moment $q$ for AMPT (SM) generated data. The plot shows the increasing trend of $d_{q}$ with $q$. The increasing trend of values of $d_{q}$ with $q$ indicates many things:

(i) Fluctuations in higher order: As the order of the moment increases, the contributions of intense fluctuations become more prominent \& significant; this leads to a higher anomalous dimension.

(ii) Multifractal nature: increasing trend of $d_{q}$ with $q$ indicates that the fluctuations are non-uniformly distributed.

(iii) Critical phenomena: slight change in order parameter leading to large-scale fluctuation and leading to rise in the anomalous fractal dimension \cite{dq_p1,dq_p2,dq_p3,dq_p4,dq_p5}.\par


The plot of $\langle F_q \rangle$ versus $lnM^{2}$ at 200 GeV, shown in Fig. 7, shows a strikingly similar behavior to that seen at lower energies. Interestingly, the values of $\langle F_q \rangle$ for different $q's$ at 200 GeV are very much similar to those at 10 AGeV energy, showing a clear and consistent pattern at a FAIR energy \& at a RHIC energy.

\section{Summary}
\label{sec:4}

The AMPT (default) focuses mainly on hadronic interactions with a limited partonic phase, while the AMPT (SM) assumes a full partonic phase and is better suited for heavy ion collisions where a quark-gluon plasma is expected to form. Also, string melting mode of AMPT might be suitable for strongly interacting matter under extreme dense conditions in HIC.

In this article, the SFM is analyzed for AMPT-generated data at 10 AGeV Au+Au collisions. The study has been done in two-dimensional $(P_{x}-P_{y})$ space. It is found that the AMPT-generated data with both modes shows some noticeable signature of intermittency. Thus, the multiparticle correlations that could be observed in AMPT data for charged particles produced in HIC.

From the present investigation, it has been observed that the intermittency indices for $q=$5 and 6 are significantly higher in the AMPT (String Melting) generated data compared to the AMPT (Default) data. This pronounced increase in intermittency indices suggests that the string melting mode, with its more extensive partonic interactions and richer dynamics, leads to stronger and more complex dynamical fluctuations. These findings highlight the string melting mode's heightened sensitivity to the underlying physics, potentially offering deeper insights into the behavior of the quark-gluon plasma and related critical phenomena.

\section*{Acknowledgements}
\label{sec:5}
The authors thankfully acknowledge the AMPT group for publicly providing the codes that have been used for generating events for this work. Authors also would like to thank Mr. Subir Mandal for his help to generate the data. The authors also acknowledge the Department of Science and Technology (DST) and the Department of Atomic Energy, Government of India, for providing funds through Project No. SR/MF/PS-02/2010 (E-6133) dated 08/10/2021.

\end{document}